\documentclass[prb,amsmath,amssymb,floatfix,twocolumn,groupedaddress]{revtex4-1}
\usepackage{graphicx}
\usepackage{hyperref}
\usepackage{enumerate}
\renewcommand{\vec}[1]{\mathbf{#1}}

\begin{document}

\title{Pairing of particle-hole symmetric composite fermions in half-filled Landau level}

\author{Zhiqiang Wang}
\affiliation{Department of Physics and Astronomy, University of
California Los Angeles, Los Angeles, California 90095-1547}
\author{Sudip Chakravarty}
\affiliation{Department of Physics and Astronomy, University of
California Los Angeles, Los Angeles, California 90095-1547}

\date{\today}

\begin{abstract}
  In a recent proposal  of the half-filled Landau level, the composite fermions are taken to be Dirac particles and particle-hole symmetric. Cooper pairing of these composite fermions in different angular momentum channels, $\ell$, can give rise to different kinds of Pfaffian states. In addition to the well known Moore-Read Pfaffian and anti-Pfaffian states, a new putative particle-hole symmetric Pfaffian state, corresponding to the $s-$wave pairing channel, was also proposed. However, the possible underlying pairing mechanism is not clear at all. In this work we provide a specific pairing mechanism for realizing some of these Pfaffian states. We show that there can be nonzero pairing in angular momentum channels $|\ell|\ge 1$ depending on the magnitude of a coupling constant. There is a quantum phase transition from the Dirac composite fermi liquid state to Cooper pairing states in angular momentum channels $|\ell|\ge 1$ as the coupling constant is tuned across its critical point value. Surprisingly the particle-hole symmetric $\ell=0$ channel pairing turns out to be impossible irrespective of the size of the coupling constant.
\end{abstract}


\maketitle
\section{Introduction} \label{sec:intro}
The nature of compressible states at $\nu=1/2$ filling fraction in the fractional quantum Hall regime has been a fascinating topic since the pioneering work by Halperin, Lee, and Read (HLR)~\cite{Halperin1993}. In the HLR picture this state is interpreted as a liquid of non-relativisitc composite fermions(CFs) coupled to a fluctuating Chern-Simons (CS) gauge field. It provides a nice explanation for the experimental observations of acoustic wave propagation~\cite{Willett1990} and has been further supported by other measurements~\cite{Du1993,Willett1997}.

However, the HLR theory has a long-standing issue in that it is incompatible with the particle-hole symmetry defined for a single Landau level~\cite{Kivelson1997,Lee1998} in the zero Landau level mixing limit. According to Ref.~\onlinecite{Kivelson1997}, the particle-hole symmetry requires the Hall conductivity of electrons in a half filled Landau level to be exactly $\sigma_{xy}=\frac{1}{2}\frac{e^2}{h}$. From this it can be deduced that the Hall conductivity of the composite fermion liquid needs to be equal to $\sigma_{xy}^{(cf)}=-\frac{1}{2}\frac{e^2}{h}$. This contradicts the HLR picture where the composite fermions feel zero magnetic field at mean field level and therefore $\sigma_{xy}^{(cf)}=0$. Considering fluctuations of the CS gauge fields beyond the mean field level can lead to a nonzero $\sigma_{xy}^{(cf)}$; however, it is small and can not explain the large value of $-\frac{1}{2}\frac{e^2}{h}$. Experimentally the measured $\sigma_{xy}$  is very close to $\frac{1}{2}\frac{e^2}{h}$~\cite{Wong1996}, and the emergent particle-hole symmetry has also been supported by several experiments~\cite{Shahar1995,Shahar1996,Kamburov2014,Kamburov2014a}. Therefore these experimental results contradict the HLR picture, which apparently breaks the particle-hole symmetry.

To resolve this issue recently Son~\cite{Son2015} has proposed a particle-hole symmetric theory in which the underlying composite fermions are taken to be Dirac particles. As in the HLR picture, these Dirac CFs are electrically neutral and coupled to an emergent gauge field. However, in this new picture the emergent gauge field does not have a Chern-Simons term. This proposal has sparked a great deal of interest~\cite{Balram2015,Barkeshli2015,Geraedts2016,Kachru2015,Murthy2016,Wang2016,Metlitski2016,Potter2016,Wang2016b,Mulligan2016,Zucker2016,Milovanovic2016} because it can not only resolve the old particle-hole symmetry issue but also provide another avenue to study some seemingly completely unrelated topics such as strongly-correlated topological insulator surface states~\cite{Wang2016}.

Cooper pairing of the CFs  gives rise to incompressible gapped states, which can possess nonabelian braiding statistics~\cite{Moore1991,Read2000} and are potentially useful for topological quantum computation~\cite{Nayak2008}. In the traditional HLR picture, pairing CFs can lead to the Moore-Read Pfaffian state~\cite{Moore1991}. However, because of the lack of the particle-hole symmetry, the anti-Pfaffian state~\cite{Lee2007,Levin2007} \textemdash the particle-hole conjugate of the Moore-Read Pfaffian state \textemdash is impossible. In contrast, in the new Dirac CF theory both the Moore-Read Pfaffian and anti-Pfaffian states are put on equal footing. Furthermore, a new putative particle-hole symmetric pairing state~\cite{Son2015} has also been proposed. However, the possible underlying pairing mechanism for realizing these states is not clear at all. In fact, in a recent work~\cite{Kachru2015} the authors there considered an effective interaction, derived from the original electron-electron Coulomb interaction, in the BCS pairing channel and found that there is no pairing instability in any angular momentum channel $\ell$.

Here we pursue an idea proposed in  Ref~\onlinecite{Chung2013} and applied to the HLR problem~\cite{Wang2014} by our group previously. We construct a specific paring mechanism for the new Dirac CF theory. This mechanism is similar to Kohn-Luttinger's mechanism~\cite{Kohn1965} in the sense that in both cases the instability results from a repulsive force. However, they are different in two respects: (1) in our mechanism a finite coupling constant is required to achieve the pairing state, while in Kohn-Luttinger's case even an infinitesimal interaction can trigger the pairing instability; (2)our new pairing mechanism takes advantage of the attraction in Matsubara frequency domain due to a dynamic screening from the finite density of fermions; while Kohn-Luttinger's idea is to take advantage of the spatial part of attraction due to a static screening.

Applying the mechanism of dynamic screening, we show that there can be nonzero pairing of Dirac CFs in angular momentum channels $|\ell| \ge 1$, including the Moore-Read Pfaffian and the anti-Pfaffian states, which correspond to the $\ell=\mp 2$ channels. There is a quantum phase transition from the Dirac CF liquid state to the $|\ell|\ge 1$ pairing states as we tune the effective coupling constant. However, the $\ell=0$ channel pairing, which is particle-hole symmetric and corresponds to the putative particle-hole symmetric Pfaffian state proposed in Ref.~\onlinecite{Son2015} turns out to be impossible in our current treatment.

The outline of our paper is as follows. In Sec.~\ref{sec:model} we define the Dirac CF model, discuss the bare propagator we use for the emergent gauge field, and obtain the dynamic screened interaction between the Dirac CFs. Then the interaction is used to solve the pairing gap equation numerically and the results are presented in Sec.~\ref{sec:result}. In Sec.~\ref{sec:discussion} we discuss the implications of our numerical results for the spatial angular momentum pairing channels of the Dirac CFs. A brief summary of our conclusion is included in Sec.~\ref{sec:conclusion}. Appendix~\ref{sec:Vbcs} and \ref{sec:RT} contain derivations of Eq.~\eqref{eq:Sint2} of Sec.~\ref{sec:model} and the current-current response function of the Dirac CF liquid within the random phase approximation.

\section{Model} \label{sec:model}
The low energy effective(Euclidean) action of the Dirac CF field $\psi$ in $(2+1)$ dimension is~\cite{Son2015}
\begin{gather}
S_{\mathrm{CF}}=\int d\tau d^2x  \lbrace \bar{\psi}  \gamma_{\mu}( \partial_{\mu} + i a_{\mu})\psi -i\frac{B}{4\pi} a_0 \rbrace, \label{eq:action}
\end{gather}
where $\bar{\psi}=\psi^{\dagger}\gamma_0$ and the two-component  $\psi$  carries  pseudo-spin degrees of freedom. The three gamma matrices
are chosen to be $ \{\gamma_0,\gamma_1,\gamma_2\} \equiv \{\sigma_3,\sigma_1,\sigma_2\}$ with $\sigma_1,\sigma_2,\sigma_3$ the three Pauli matrices. We have also set $\hbar=e=c=1$ and the Fermi velocity $v_F=1$. $a_{\mu}$ is the emergent gauge field that the Dirac CFs couple to. $\mu=0,1,2$ represent the imaginary time $\tau$ and the spatial $x,y$ variables. Throughout the whole paper we will use Greek  subscripts to denote the time-spatial $3-$vector components and Latin  subscripts for the spatial vectors. In the 2nd term of Eq.~\eqref{eq:action}, $B$ is the  physical magnetic field. As mentioned before, the key differences between $S_{\mathrm{CF}}$ and the HLR theory are: there is no Chern-Simons term for the gauge field $a_{\mu}$, and the new CFs are relativistic particles. Differentiating $S_{\mathrm{CF}}$ with respect to $a_0$ gives $\bar{\psi}\gamma^0 \psi= \frac{B}{4\pi}$, which shows that the Dirac CF density is equal to twice  the flux density, B. Therefore to electrons, the Dirac CFs act as  double vortex objects. Furthermore the emergent magnetic field strength is given by $b(\vec{x})=\nabla \times \vec{a}=4\pi \rho_e^{\prime}(\vec{x})$. Here $\rho_e^{\prime}=\rho_{e}-\rho_{\nu=1/2}$ is the deviation of the original electric charge density $\rho_e$ from its half filled Landau level value $\rho_{\nu=1/2}$. A prime is added to $\rho_e^{\prime}$ to distinguish it from $\rho_e$. We see that the original electric field and magnetic flux density interchange their roles in the  Dirac CF theory. In this sense the new theory is electro-magnetic dual~\cite{Kachru2015} to the original electron problem.

At the mean field level $<b(\vec{x})>=\rho^{\prime}_e=0$ because the electron Landau level is exactly at half filling. Therefore the Dirac CFs are described by a free Dirac Hamiltonian $H= v_F\vec{p}\cdot(\hat{z}\times \boldsymbol{\sigma})$ at zero field $b=0$. The dispersion is simply a Diarc cone with two branches $\epsilon_{\vec{k}}^s=s\,\hbar v_F |\vec{k}|$, where $s=\pm$. Because the Dirac CF has a finite density $B/4\pi$, the Fermi energy $\epsilon_F=v_F k_F$ is finite with the Fermi wavevector given by $k_F=\sqrt{B}$, same as that in the HLR picture.

Beyond the mean field level there will be gauge field fluctuations of $a_{\mu}$. In general these fluctuations can mediate some effective interaction and may provide the necessary pairing glue between the Dirac CFs. To find out that answer we need one more ingredient: the dynamics of the gauge fields. This could be given by an emergent Maxwell term in the action: $ S_{\mathrm{Max}}\propto -\frac{1}{4 g^2} \int d^3x f_{\mu\nu}^2 $,  where $f_{\mu\nu}=\partial_{\mu} a_{\nu}-\partial_{\nu}a_{\mu}$ is the field strength and $g$ is the coupling constant. The other choice is to use the original Coulomb interaction between electrons: $\frac{e^2}{\epsilon_r |\vec{x}-\vec{x}^{\prime}|}\rho_e^{\prime}(\vec{x})\rho_e^{\prime}(\vec{x}^{\prime})$, where $\epsilon_r$ is the background dielectric constant, and translate it into an interaction between the emergent gauge fields $a_{\mu}$ by using $b(\vec{x})=4\pi \rho_e^{\prime}(\vec{x})$. Written in the momentum space, the action of this term reads~\cite{Kachru2015}
\begin{align}
S_{\mathrm{Coulomb}}  =& \frac{1}{2} \int \frac{d\Omega d^2 \vec{q}}{(2\pi)^3} \; a_{T}(\Omega,\vec{q})  \frac{2\pi e^2}{\epsilon_r |\vec{q}|} \frac{|\vec{q}|^2}{16\pi^2} \; a_{T}(-\Omega,-\vec{q} ). \label{eq:Scoulomb}
\end{align}
The frequency $\Omega$ here and elsewhere should be understood as Matsubara frequencies. The temperature has been already set to $T=0$ so that all Matsubara frequencies are continuous. In the action $S_{\mathrm{Coulomb}}$ only the spatial transverse component of gauge fields is involved, as indicated by the subscript ``T" in $a_{\mathrm{T}}(\Omega,\vec{q})=\epsilon_{ij} \hat{q}_i a_j(\Omega,\vec{q})$, where $\epsilon_{ij}$ is the antisymmetric tensor and the summation convention is assumed. In Eq.~\eqref{eq:Scoulomb} the $2\pi e^2/\epsilon_r |\vec{q}|$ factor comes from the 2D Fourier transform of the electron-electron Coulomb interaction. The factor $\frac{|\vec{q}|^2}{16\pi^2}$  comes from the conversion from $\rho_e^\prime$ to $a_{\mathrm{T}}$. Now we see that the Lagrangian density of $S_{\mathrm{Coulomb}}$ is $\propto |\vec{q}|$ while that of the Maxwell term $S_{\mathrm{Max}}$ is $\propto f^2_{\mu\nu} \propto |\vec{q}|^2$. Therefore in the long wavelength limit, the Coulomb term dominates and the Maxwell term~\cite{Kachru2015} can be dropped. Taking $S_{\mathrm{Coulomb}}$ as our bare gauge field action we can readily read off the bare gauge field propagator inverse, which has only the transverse component
\begin{gather}
[ \mathcal{D}_{\mathrm{T}}^{(0)}]^{-1}(\Omega,\vec{q})= \frac{e^2}{8 \pi \epsilon_r} |\vec{q}|. \label{eq:prop-bare}
\end{gather}
We add the superscript  in $\mathcal{D}_{\mathrm{T}}^{(0)}$ to indicate that it is the bare gauge field propagator without the screening from the finite density Dirac CFs.

Integrating out the gauge fields $a_{\mathrm{T}}$ gives a current-current interaction between Dirac CFs described by the following action~\cite{Kachru2015}
\begin{align}
S_{\mathrm{int}} & = \frac{1}{2} \int \frac{ d\Omega d^2 \vec{q}}{(2\pi)^3} J_{\mathrm{T}}(\Omega,\vec{q})  \mathcal{D}_{{\mathrm{T}}}^{(0)}(\Omega,\vec{q}) J_{{\mathrm{T}}}(-\Omega,-\vec{q}),   \label{eq:Sint}
\end{align}
where $J_{{\mathrm{T}}}(\Omega,\vec{q})= \epsilon_{ij} \hat{q}_i J_{j}(\Omega,\vec{q})$ is the transverse component of CF current operator. For the Dirac CF the current density $J_{i}(\Omega,\vec{q})= v_F\int d\omega d^2 \vec{k}/(2\pi)^3 \psi^{\dagger}(\omega+\Omega,\vec{k}+\vec{q})\, i \gamma_0\gamma_i \psi(\omega,\vec{k})$ is equivalent to the transverse pseudo-spin density. Because of $\gamma_0\gamma_i$, $J_i$ mixes particles near the CF Fermi surface with their anti-particles  buried deep in the Dirac sea. However, our theory is trustworthy only as a low energy effective theory near the Fermi surface. Therefore we need to project  the interaction  near the Fermi surface. This can be done by replacing $\psi(\omega,\vec{k})$ by~\cite{Kachru2015}
\begin{gather}
P^{(+)}_{\vec{k}} \psi(\omega,\vec{k})=\frac{1}{\sqrt{2}} \begin{pmatrix} i e^{-i\theta_{\vec{k}}} \\ 1  \end{pmatrix} \chi(\omega,\vec{k}) \label{eq:ppluspsi}
\end{gather}
in the definition of $J_{\mathrm{T}}(\Omega,\vec{q})$. Here $P^{(+)}_{\vec{k}} \equiv \frac{1}{2}[1+ i \gamma_0 \boldsymbol{\gamma} \cdot \hat{\vec{k}}]$ is the projection operator for the positive energy branch $\epsilon_{\vec{k}}^{+}$, for the low energy qausiparticles near the Fermi surface. $\theta_{\vec{k}}$ is the azimuthal angle of the momentum vector $\vec{k}$ in the $xy$ plane. $\chi(\omega,\vec{k})$ is now a scalar field representing quasiparticles near the Fermi surface. Substituting $\psi(\omega,\vec{k})$ in the Eq.~\eqref{eq:Sint} with $P^{(+)}_{\vec{k}} \psi(\omega,\vec{k})$ leads to(for details see Appendix~\ref{sec:Vbcs} and Ref.~\onlinecite{Kachru2015})
\begin{align}
S_{\mathrm{int}}     & = \frac{1}{2} \int \prod_{i=1}^4 \frac{d^3 k_i}{(2\pi)^3}  \; (2\pi)^3 \delta^{(3)}(k_3+k_4-k_1-k_2) \nonumber \\
    & \quad \times \frac{8\pi v_F^2 \epsilon_r}{e^2} \frac{\exp\{ -\frac{i}{2}[\theta_{\vec{k}_1}+\theta_{\vec{k}_2}-\theta_{\vec{k}_3}-\theta_{\vec{k}_4}]\}}{|\vec{k}_1-\vec{k}_3|} \nonumber \\
    &\quad \times \chi^{\dagger}(k_4)\chi(k_2) \chi^{\dagger}(k_3) \chi(k_1), \label{eq:Sint2}
\end{align}
where for brevity we have adopted the relativistic  notation $k_i=(\omega_i,\vec{k}_i)$. The magnitude of the vector $\vec{k}_i$ will be denoted as $|\vec{k}_i|$ to avoid any confusion.

Then consider the above interaction in the BCS channel: $k_1=-k_2 =k \equiv(\omega,\vec{k})$, and $k_3=-k_4 =k^{\prime} \equiv(\omega^{\prime},\vec{k}^{\prime})$, and introduce the frequency and momentum transfer as: $\Omega=\omega^\prime -\omega $ and $\vec{q}=\vec{k}^{\prime}-\vec{k}$. We make Fermi surface approximations: $|\vec{k}|=|\vec{k}^{\prime}|=k_F$. Then the interaction in the BCS channel can be readily read  from $S_{\mathrm{int}}$ as~\cite{Kachru2015}
\begin{gather}
V_{\mathrm{BCS}}(\vec{k}^\prime,\vec{k})= \frac{8\pi v_F^2 \epsilon_r}{2 k_F e^2} \frac{e^{-i[\theta_{\vec{k}}-\theta_{\vec{k}^\prime}]}}{|\sin\frac{\theta_{\vec{k}}-\theta_{\vec{k}^\prime}}{2}|}.
\end{gather}
As has been pointed out in Ref.~\onlinecite{Kachru2015}, this interaction is repulsive in all angular momentum channels and therefore can not give rise to any pairing. Notice that the phase factor in the numerator comes from the difference of the Berry phase carried by a Cooper pair, which is made of a spinor $P^{(+)}_{\vec{k}} \psi(\omega,\vec{k})$ and another spinor with momentum $ -\vec{k}$, from that carried by another Cooper pair with momenta:$\{\vec{k}^\prime,-\vec{k}^\prime\}$.

Now we incorporate the screening effects from the finite density Dirac CFs on the gauge field within the random phase approximation(RPA). In the limit $|\Omega| <v_F |\vec{q}|\ll \epsilon_F$, the transverse current-current response function $\mathcal{R}_{\mathrm{T}}$ of the Dirac CFs is given by $ \mathcal{R}_{\mathrm{T}}(\Omega,\vec{q}) = - \frac{\epsilon_F}{2\pi } \frac{|\Omega|}{v_F |\vec{q}|}$ (for details see Appendix~\ref{sec:RT}) and the RPA renormalized gauge field propagator can be obtained from the Dyson equation
\begin{gather}
\mathcal{D}_{\mathrm{T}}^{\mathrm{RPA}}(\Omega,\vec{q})= \frac{\mathcal{D}_{\mathrm{T}}^{(0)}(\Omega,\vec{q})}{1 -\mathcal{R}_{\mathrm{T}}(\Omega,\vec{q})\mathcal{D}_{\mathrm{T}}^{(0)}(\Omega,\vec{q})}.
\end{gather}
We can define a dynamic dielectric function as $\epsilon(\Omega,\vec{q})= 1-\mathcal{R}_{\mathrm{T}}(\Omega,\vec{q})\mathcal{D}_{\mathrm{T}}^{(0)}(\Omega,\vec{q})$ so that  $\mathcal{D}_{\mathrm{T}}^{\mathrm{RPA}}(\Omega,\vec{q})=\mathcal{D}_{\mathrm{T}}^{(0)}(\Omega,\vec{q})/\epsilon(\Omega,\vec{q})$. Similar to $\mathcal{D}_{\mathrm{T}}^{(0)}$, $\mathcal{D}_{\mathrm{T}}^{\mathrm{RPA}}$ will mediate a current-current interaction between Dirac CFs, which is now frequency dependent and given by $V_{\mathrm{eff}}(\Omega;\vec{k}^\prime,\vec{k})=V_{\mathrm{BCS}}(\vec{k}^\prime,\vec{k})/\epsilon(\Omega,\vec{q})$. The interaction $V_{\mathrm{eff}}$ can have considerable attraction at higher frequencies in certain angular momentum channels, as will be seen clearly in Sec.~\ref{sec:result}, and therefore can lead to a net pairing for the scalar fermion field $\chi(k)$.

Because $\chi(k)$ and the projected Dirac spinor $P^{(+)}_{\vec{k}}\psi(k)$ are connected by Eq.~\eqref{eq:ppluspsi}, pairing of $\chi(k)$ also indicates pairing of $P^{(+)}_{\vec{k}}\psi(k)$. However, their spatial  angular momentum pairing channels can be different because of the nontrivial Berry phase factor carried by the spinor $P^{(+)}_{\vec{k}}\psi(k)$ on the right hand side of Eq.~\eqref{eq:ppluspsi}. As ultimately we are interested in the pairing of $P^{(+)}_{\vec{k}}\psi(k)$, we use $\ell$ to denote its angular momentum channel, while for $\chi(k)$, we use $\ell^\prime$: $<\chi(-k)\chi(k)> \propto \Delta_{\ell^\prime}(\omega) e^{i\ell^{\prime} \theta_{\vec{k}}}$, where $\Delta_{\ell^\prime}(\omega)$ is the frequency dependent pairing gap of that channel. Here we have assumed that: (1) the dependence of the order parameter on the frequency $\omega$ and on the momentum direction $\theta_{\vec{k}}$ along the Fermi surface can be separated; (2) pairing of different angular momentum channels $\ell^\prime$ are decoupled from each other. The relationship between $\ell$ and $\ell^\prime$ depends on whether or not the pairing order parameter of $P^{(+)}_{\vec{k}}\psi(k)$ is in a singlet channel or a triplet channel in the pseudo-spin space, as will be discussed in detail later on in Sec.~\ref{sec:discussion}.

We first discuss pairing in terms of $\chi(k)$. Integrating $V_{\mathrm{eff}}(\Omega;\vec{k}^\prime,\vec{k})$ over the Fermi surface in the $\ell^{\prime}$ channel defines a dimensionless effective interaction:
\begin{align}
\widetilde{V}_{\ell^\prime}(\widetilde{\Omega}) & \equiv N_0 \int\frac{d [\theta_{\vec{k}}-\theta_{\vec{k}^\prime}]}{2\pi} V_{\mathrm{eff}}(\Omega;\vec{k}^\prime,\vec{k}) e^{i\ell^{\prime} [\theta_{\vec{k}}-\theta_{\vec{k}^\prime}]}  \\
& = \alpha \int_{-\pi}^{\pi} \frac{d\theta}{2\pi}   \frac{e^{i[\ell^\prime-1] \theta}}{|\sin\frac{\theta}{2}|} \frac{2}{1+\alpha \frac{|\widetilde{\Omega}|}{\sin^2 \frac{\theta}{2}}}, \label{eq:Vtilde}
\end{align}
where $|\widetilde{\Omega}|=|\Omega|/\epsilon_F$. $N_0=\epsilon_F/2\pi v_F^2$ is the density of states of the free Dirac CFs at the Fermi energy $\epsilon_F$. In the above equation we have introduced an effective coupling constant $\alpha=N_0 4\pi v_F^2 \epsilon_r/2 k_F e^2=1/\alpha^{\prime}$, with $\alpha^{\prime}\equiv e^2/\epsilon_r v_F$ the fine structure constant of the Dirac CFs. That it is $1/\alpha^\prime$, instead of $\alpha^\prime$, serves as our coupling constant reflecting the electromagnetic duality~\cite{Kachru2015} in our problem. Since we do not have a good estimation of  $v_F$ for the composite fermions, in the following we treat $\alpha$ as a generic tunning parameter. In Sec.~\ref{sec:result} we will show that  $\alpha$ governs a quantum phase transition from a Dirac composite Fermi liquid state to a pairing state of the composite fermions. Note that $\alpha$ can be also rewritten as $\alpha=\frac{v_F k_F}{e^2 k_F/\epsilon_r}$, which is nothing but the ratio between the kinetic energy and the Coulomb interaction energy at the length scale $k_F^{-1}$. This shows that the driving force behind our phase transition is a competition between the dynamic screening from the finite density composite fermions and the bare static repulsive Coulomb interaction, because the kinetic energy $v_F k_F=\epsilon_F$ characterizes the strength of the dynamic screening as we can see from the expression of the response function $\mathcal{R}_{\mathrm{T}}$.

With $\widetilde{V}_{\ell^\prime}$ we can solve the self-consistent equation of the pairing gap $\Delta_{\ell^\prime}(\omega)$, which is given by~\cite{Chung2013}
\begin{gather}
\widetilde{\Delta}_{\ell^\prime}(\widetilde{\omega})=-\int \frac{d\widetilde{\omega}^{\prime}}{2\pi} \;  \widetilde{V}_{\ell^\prime}(\widetilde{\omega}-\widetilde{\omega}^\prime) \frac{\widetilde{\Delta}_{\ell^\prime}(\widetilde{\omega}^{\prime})}{\sqrt{(\widetilde{\omega}^\prime)^2+|\widetilde{\Delta}_{\ell^\prime}(\widetilde{\omega}^\prime)|^2}}, \label{eq:deltatilde}
\end{gather}
where all quantities with tilde are dimensionless: frequencies $\widetilde{\omega}=\omega/\epsilon_F,\widetilde{\omega}^\prime=\omega^\prime/\epsilon_F$ and the pairing gap $\widetilde{\Delta}_{\ell^\prime}(\widetilde{\omega})=\Delta_{\ell^\prime}(\omega)/\epsilon_F$. As emphasized in our previous work~\cite{Chung2013,Wang2014}, a complete solution to the paring problem needs to take into account the fermion wavefunction renormalization factor $Z(\omega)$. This factor has an anomaly on the Fermi surface: $Z(\omega) \rightarrow 0$ as $\omega \rightarrow 0$ in the Dirac CF liquid phase, similar to the HLR picture. However, in a pairing phase this anomaly will be cutoff at frequencies of the order of the pairing gap. In other words, setting $Z(\omega)\approx 1$ should be qualitatively correct as long as the coupling constant $\alpha$ is not too close to its critical point value $\alpha_c$.

In the next section we first plot out $\widetilde{V}_{\ell^\prime}(\widetilde{\Omega})$ to show that $\widetilde{V}_{\ell^\prime}$ has a sizable attraction at high frequencies for $|\ell^\prime -1| \ge 1$; while it is repulsive in the entire frequency range for $\ell^\prime=1$. Then we present our numerical results to the above self-consistent gap Equation~\eqref{eq:deltatilde} for different channels $\ell^\prime$.

\section{Results}\label{sec:result}
\subsection{The effective interaction $\widetilde{V}_{\ell^\prime}(\widetilde{\Omega})$}
\begin{figure}[htp]
 \centering
 \includegraphics[width=\linewidth]{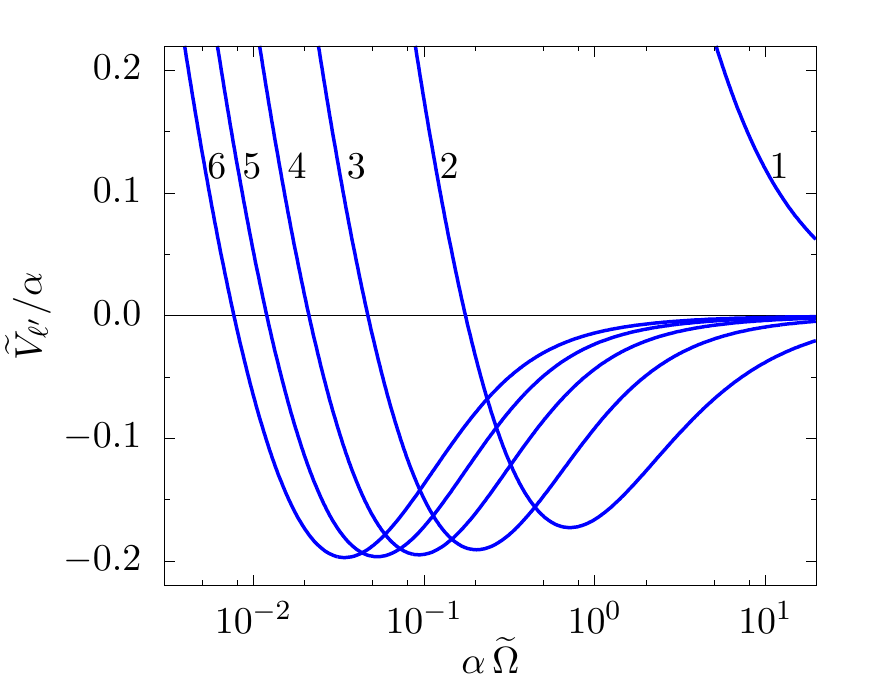}
 \caption{Loglinear plots of $\widetilde{V}_{\ell^\prime}(\widetilde{\Omega})/\alpha$ as a function of $\alpha \, \widetilde{\Omega}$ for $\ell^\prime=1,2,3,4,5,6$. Each curve is labeled by its $\ell^\prime$ value. Notice that the horizontal axis is in a logarithmic scale. }
 \label{fig:fig1}
\end{figure}

We first notice that $\widetilde{V}_{\ell^\prime}(\widetilde{\Omega})$ only depends on the magnitude of $\ell^\prime-1$ but not its sign. Therefore we only need to plot  $\widetilde{V}_{\ell^\prime}(\widetilde{\Omega})$ for channels $\ell^\prime -1 \ge 0$. This is shown in Fig.~\ref{fig:fig1}.
From this plot  we see that the effective interaction in the $\ell^\prime=1$ channel is completely repulsive in the entire frequency range. Therefore there can not be any pairing in this channel.

For $|\ell^\prime-1| \ge 1$, although $\widetilde{V}_{\ell^\prime}(\widetilde{\Omega})$ is repulsive and diverges logarithmically as $\widetilde{V}_{\ell^\prime}(\widetilde{\Omega}) \sim -\frac{\alpha}{\pi} \log |\widetilde{\Omega}|$ in the small frequency limit $\widetilde{\Omega} \rightarrow 0$,  it has considerable attraction at higher frequencies. The balance between the repulsion and attraction is controlled by the coupling constant $\alpha$. Depending on the magnitude of $\alpha$, such a balance may be tipped  in favor of a pairing state. In the next subsection we show that this is indeed the case by solving the gap Equation~\eqref{eq:deltatilde} numerically.

\subsection{Solution to the pairing gap equation} \label{sec:solution}
We first cut off the effective interaction $\widetilde{V}_{\ell^\prime}(\widetilde{\Omega})$ and the frequency integral at $|\widetilde{\Omega}|\ge 1$ in the gap Eq.~\eqref{eq:deltatilde}. We choose the frequency grid to be $5\times 10^4$ and solve the gap equation by iteration until a desired convergence accuracy is achieved, following Ref.~\onlinecite{Chung2013}.

\begin{figure}[htp]
\centering
 \includegraphics[width=\linewidth]{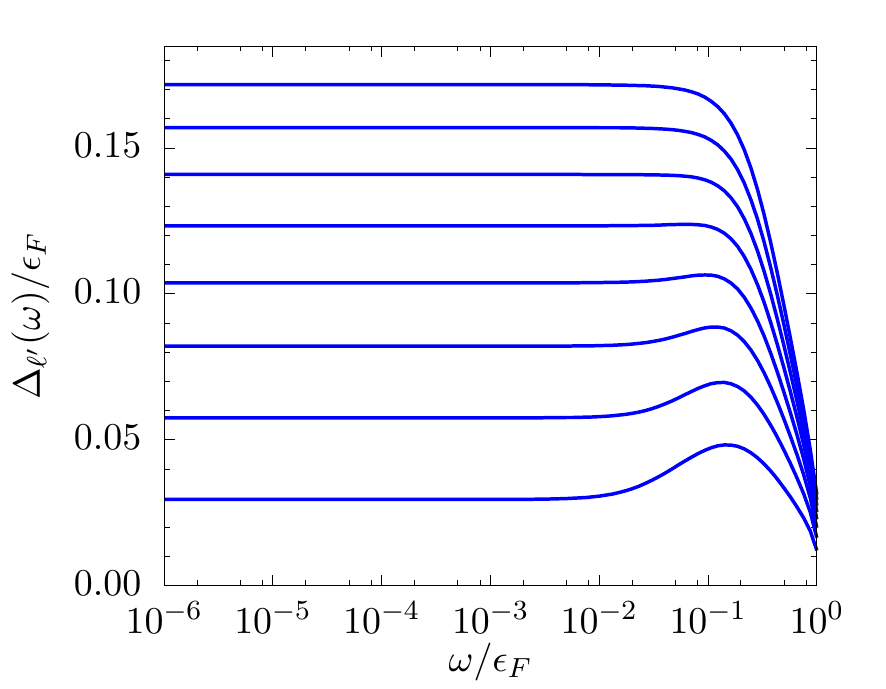}
  \caption{The plot of gap $\Delta_{\ell^\prime}(\omega)$ as a function of the frequency $\omega$ for angular momentum channel $\ell^\prime=2$. From top to bottom: $\alpha=24,22,20,18,16,14,12,10$. }
\label{fig:fig2}
\end{figure}

\begin{figure}[htp]
\centering
  \includegraphics[width=\linewidth]{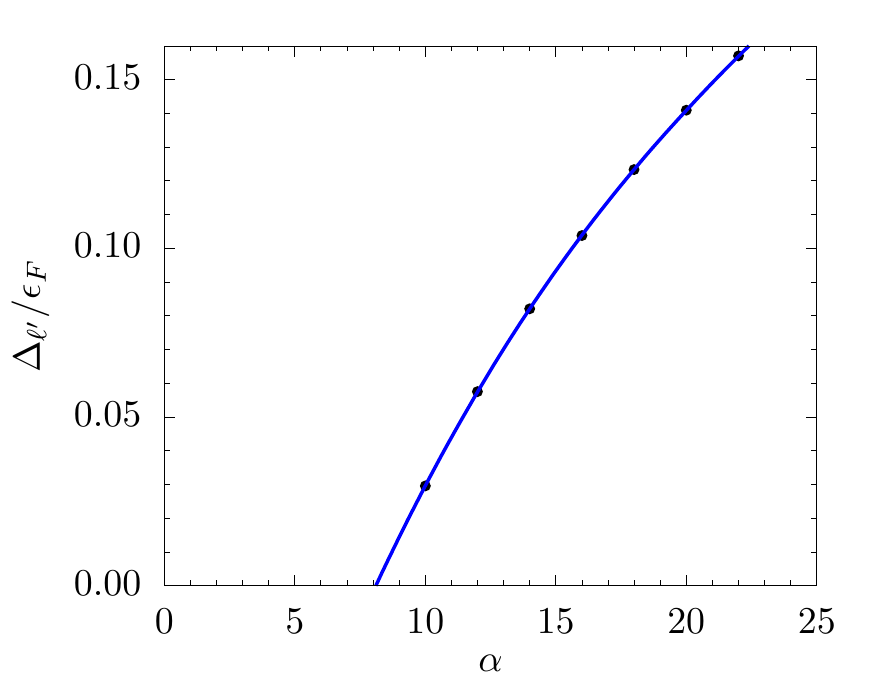}
  \caption{Plot of gap $\Delta_{\ell^\prime} =\Delta_{\ell^\prime}(\omega=0)$ as a function of $\alpha$ for angular momentum channel $\ell^\prime=2$. The critical value of  $\alpha$ is estimated to be $\alpha_c\approx 8$.}
  \label{fig:fig3}
\end{figure}

Fig.~\ref{fig:fig2} shows the results of $\Delta_{\ell^\prime}(\omega)$ for angular momentum channel $\ell^\prime=2$. Different curves in this figure correspond to different values of the coupling constant $\alpha$. From these curves we see that $\Delta_{\ell^\prime}(\omega)$ falls off rapidly as $\omega/\epsilon_F \rightarrow 1$. This is due to our sharp cutoff of $\widetilde{V}_{\ell^\prime}$ at $\omega/\epsilon_F=1$. As $\omega\rightarrow 0$, clearly $\Delta_{\ell^\prime}(\omega)$ reaches a finite constant $\Delta_{\ell^\prime}\equiv \Delta_{\ell^\prime}(\omega=0)$. Remember that we are working in Matsubara frequency space. Therefore $\Delta_{\ell^\prime}$ gives the $T=0$ thermodynamic pairing gap. As we can see in Fig.~\ref{fig:fig2} the size of $\Delta_{\ell^\prime}$ decreases progressively as we decrease the coupling constant from $\alpha=24$ to $\alpha=10$. Naturally we would expect that $\Delta_{\ell^\prime}$ eventually vanishes at certain critical value of $\alpha=\alpha_c$, below which the Dirac CF liquid state is stabilized instead. This is explicitly shown for $\ell^\prime=2$ in Fig.~\ref{fig:fig3}, where we plot  $\Delta_{\ell^\prime}$  extracted from Fig.~\ref{fig:fig2} for different values of $\alpha$. An extrapolation to $\Delta_{\ell^\prime}=0$ shows that the gap vanishes at $\alpha_c\approx 8$. Therefore indeed there is a quantum phase transition from the $\ell^\prime=2$ pairing state to the Dirac CF liquid state as we decrease $\alpha$ across $\alpha_c$ towards zero.

\begin{figure}[ht]
  \centering
  \includegraphics[width=\linewidth]{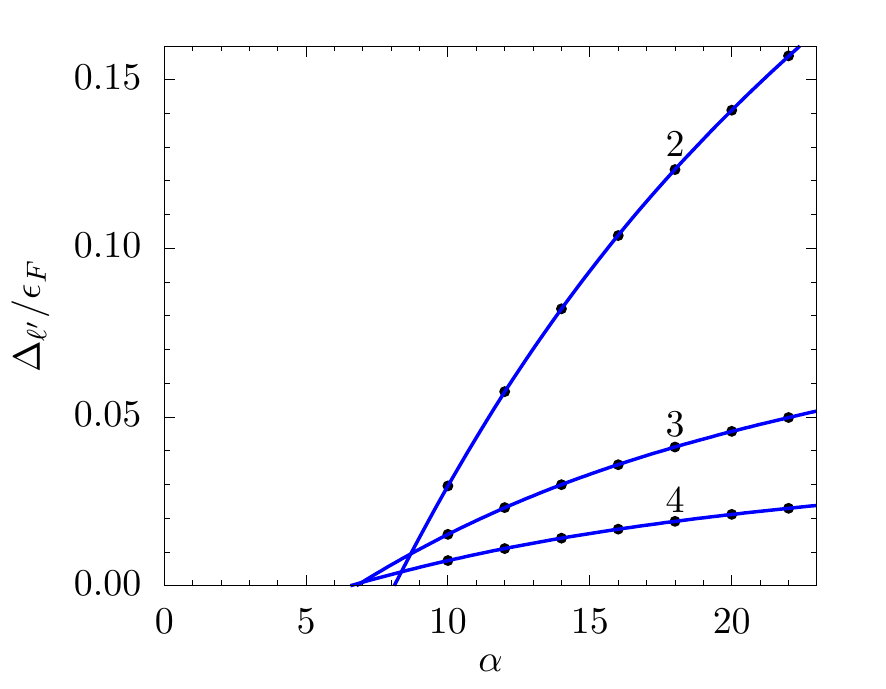}
  \caption{Plots of the gap $\Delta_{\ell^\prime}=\Delta_{\ell^\prime}(\omega=0)$ as a function of the effective coupling constant $\alpha$ for angular momentum channel $\ell^\prime=2,3,4$. The critical coupling constants for different channels are very close to each other. For $\ell^\prime=2$, $\alpha_c \approx 8$; while for $\ell^\prime \ge 3$, $\alpha_c\approx 7$. We can not resolve the difference between $\alpha_c$ for different $\ell^\prime \ge 3$ channels.}
  \label{fig:fig4}
\end{figure}

We can do similar analysis for other higher angular momentum channels $\ell^\prime$. The results for $\ell^\prime=2,3,4$ are displayed in Fig.~\ref{fig:fig4}. From this plot we see that at the same coupling constants $\Delta_{\ell^\prime}$ decreases as $\ell^\prime$ increases. This is understandable given that the frequency range of sizable attraction in $\widetilde{V}_{\ell^\prime}(\widetilde{\Omega})$ decreases with $\ell^\prime$, as we can see from Fig.~\ref{fig:fig1} \,(notice that the horizontal frequency axis is logarithmic). This feature is also similar to what have been found in the transverse gauge field problem in Ref.~\onlinecite{Chung2013}. However, unlike there the critical value $\alpha_c$ for different channels are very close to each other in our present problem. For $\ell^\prime \ge 3$, $\alpha_c\approx 7$ and we can not resolve the difference between $\alpha_c$ for different channels.

\section{Discussion}\label{sec:discussion}
As already mentioned before, the pairing angular momentum channel $\ell^{\prime}$ of the order parameter $\langle \chi(-k)\chi(k)\rangle$ is different from that of the pairing order parameter in terms of pseudo-spinors $P^{(+)}_{\vec{k}}\psi(k)$ because of the spinor's Berry phase factor. Now we discuss the pairing order parameters $\hat{\Delta}$ in terms of the $P^{(+)}_{\vec{k}}\psi(k)$ field. A hat has been put on $\hat{\Delta}$ to indicate that $\hat{\Delta}$ is a $2\times 2$ matrix. In general we can parametrize this matrix as
\begin{gather}
\hat{\Delta}(k)=[\Delta_s(k)+\vec{d}(k)\cdot \boldsymbol{\sigma}] i\sigma_2,
\end{gather}
where $\Delta_s(k)$ gives the pairing order parameter in the pseudo-spin singlet channel and the vector $\vec{d}(k)$ characterizes the triplet channel pairing. In the following discussion we assume that the Cooper pairing in these two channels are approximately decoupled when only the low energy degrees of freedom near the Fermi surface participate in the pairing.

The pseudo-spin singlet channel order parameter $\hat{\Delta}(k)=\langle \psi^{T}(-k) P^{(+)}_{-\vec{k}} i \sigma_2 P^{(+)}_{\vec{k}} \psi(k) \rangle$ is the one adopted in Ref.~\onlinecite{Son2015}. Substituting the definition of $P^{(+)}_{\vec{k}}\psi(k)$ from Eq.~\eqref{eq:ppluspsi} into this order parameter leads to $\hat{\Delta}(k) \propto i \sigma_2 \, e^{-i\theta_{\vec{k}}} \langle \chi(k) \chi (-k) \rangle$. Therefore if $<\chi(-k)\chi(k)> \propto e^{i \ell^\prime \theta_{\vec{k}}}$ is in the $\ell^\prime$ channel, then the order parameter $\hat{\Delta}(k) \propto  i\sigma_2\, e^{i[\ell^\prime-1] \theta_{\vec{k}}}$ is in the $\ell=\ell^{\prime}-1$ channel. Because Fermi statistics requires $\hat{\Delta}(k)$ to be antisymmetric and it is already antisymmetric in the pseudo-spin space,  its spatial part needs to be \emph{symmetric}; that is, only \emph{even} $\ell$, or only \emph{odd} $\ell^\prime$, channels are allowed. However, pairing in the particle-hole symmetric~\cite{Son2015} $\ell=0$ channel is impossible because it corresponds to the completely repulsive $\ell^\prime=1$ channel. Other \emph{even} $\ell\ne 0 $ channels are all possible. These include the Moore-Read Pfaffian state, corresponding to the pairing channel  $\ell=-2$, and its particle-hole conjugate, the anti-Pfaffian state~\cite{Lee2007,Levin2007}, corresponding to the $\ell=2$ channel pairing (see Ref.~\onlinecite{Son2015}). Notice that these pairing channels are different from  HLR picture, where the Moore-Read Pfaffian state corresponds to the channel of $\ell=-1$~\cite{Read2000}  and the anti-Pfaffian state corresponds to the channel of $\ell=3$~\cite{Son2015}.

If the pseudo-spin triplet order parameter $\hat{\Delta}(k)=\langle \psi^{T}(-k) P^{(+)}_{-\vec{k}}  \vec{d}\cdot\boldsymbol{\sigma} i \sigma_2 P^{(+)}_{\vec{k}} \psi(k) \rangle$ has been chosen, then $\ell$ needs to be \emph{odd}. Pairing is possible in all these \emph{odd} $\ell$ channels except the one that corresponds to the $\ell^\prime=1$ channel of Fig.~\ref{fig:fig1}. The relation between $\ell$ and $\ell^\prime$ in the pseudo-spin triplet case depends on the form of the vector $\vec{d}$ and is different from that of the singlet case. If the pseudo-spin triplet state is $\lvert\uparrow\uparrow\rangle$, then because  $\ell=\ell^\prime-2$,  $\ell=-1$ pairing is impossible; if the triplet pairing state is $\lvert\downarrow\downarrow \rangle$, then $\ell=\ell^\prime$ and $\ell=1$ channel is impossible. Notice that the other triplet state $(\lvert\uparrow\downarrow\rangle + \lvert\downarrow\uparrow\rangle)/\sqrt{2}$ pairing is impossible for the spinor in Eq.~\eqref{eq:ppluspsi} because $\hat{\Delta}$ is identically zero.

We consider that both the pseudo-spin singlet and triplet order parameters are allowed so that the spatial pairing channel $\ell$ can be either even or odd. This is in contrast to the Ref.~\onlinecite{Son2015}, where only the pseudo-spin singlet order parameter was considered such that only even $\ell$ channels were possible. Interestingly, from the previous discussions we find that $\ell^\prime$ needs to be always odd, regardless of $\ell$ being even or odd.

\section{Conclusion}\label{sec:conclusion}
To conclude we have constructed a specific pairing mechanism for the Dirac composite fermions proposed recently for the half filled Landau levels. By taking advantage of the attraction of a dynamically screened effective interaction at high Matsubara frequencies we show that there can be nonzero pairing in the angular momentum channels $|\ell|\ge 1$ at certain coupling constant values. As the coupling constant is varied, there can be a quantum phase transition from the Dirac CF liquid state to a pairing state of CFs, which should be understood as a fractional quantum Hall state of the original electrons. Apart from the well known Moore-Read Pfaffian state(singlet $\ell=-2$ channel) and its particle-hole conjugate(singlet $\ell=2$ channel): the anti-Pfaffian state, other Pfaffian states are also possible. However, in contrast to the channels $|\ell|\ge 1$, the singlet $\ell=0$ particle-hole symmetric channel, corresponding to $\ell^\prime=1$ in Fig.~\ref{fig:fig1}, effective interaction is completely repulsive in the entire frequency range considered, which therefore renders the putative particle-hole symmetric Pfaffian state impossible in our pairing mechanism. To understand the proper behavior close to $\alpha_{c}$, we should also consider the second Eliashberg equation involving $Z(\omega)$. Work in this direction is under progress.

\begin{acknowledgments}
This research was supported by funds from David S. Saxon Presidential Term Chair at UCLA. We thank S. Raghu and S. A. Kivelson for discussion.
\end{acknowledgments}

\appendix
\section{THE INTERACTION $S_{\mathrm{int}}$ of Equation~\eqref{eq:Sint2}}\label{sec:Vbcs}
In this Appendix we recapitulate how to obtain the $S_{\mathrm{int}}$ of Eq.~\eqref{eq:Sint2} from  Equation~\eqref{eq:Sint} in the main text. More details can be found in Ref.~\onlinecite{Kachru2015}.

We start with the definition of the transverse Dirac CF current
\begin{align}
J_{\mathrm{T}}(q)& =\epsilon_{ij} \hat{q}_i J_{j}(q) \\
&=  v_F \int \frac{d^3 k_1}{(2\pi)^3} \frac{d^3 k_2}{(2\pi)^3}(2\pi)^3\delta^{(3)}(k_2-k_1-q) \nonumber \\
& \quad  \times \;\psi^{\dagger}(k_2) [\epsilon_{ij} \hat{q}_i \, i \gamma_0 \gamma_j] \psi(k_1),
\end{align}
where again the relativistic notations $k_i=(\omega_i,\vec{k}_i),q\equiv(\Omega,\vec{q})$ have been used. Then replace $\psi(k_i)$ in the above with
\begin{gather}
P^{(+)}_{\vec{k}_i} \psi(k_i)=\frac{1}{\sqrt{2}} \begin{pmatrix} i e^{-i\theta_{\vec{k}_i}} \\ 1  \end{pmatrix} \chi(k_i) \label{eq:ppluspsi-a}
\end{gather}
to project out the current carried only by the low energy degrees of freedoms near the Fermi surface. This leads to
\begin{align}
J_{\mathrm{T}}(q) & =-v_F \int \frac{d^3 k_1}{(2\pi)^3} \frac{d^3 k_2}{(2\pi)^3}(2\pi)^3\delta^{(3)}(k_2-k_1-q) \nonumber \\
&\quad \times e^{i \frac{\theta_{\vec{k}_2}-\theta_{\vec{k}_1}}{2}} \chi^{\dagger}(k_2) \chi(k_1). \label{eq:Jt}
\end{align}
In this expression the $e^{i \frac{\theta_{\vec{k}_2}-\theta_{\vec{k}_1}}{2}}$ factor reflects the fact that for our Dirac CF described by the Hamiltonian
\begin{gather}
H=\hbar v_F\vec{p}\cdot(\hat{z}\times \boldsymbol{\sigma}),
\end{gather}
the transverse pseudo-spin direction $\hat{z}\times \boldsymbol{\sigma}$ is locked to the momentum direction $\vec{p}$. So there will be a nontrivial $\pi$ Berry phase picked up by $J_{\mathrm{T}}(\Omega,\vec{q})$ if the momentum $\vec{k}_2$ is traversed around $\vec{k}_1$ by $2\pi$. Remember that $J_{\mathrm{T}}(\Omega,\vec{q})$ is nothing but the transverse spin density.

A direct substitution of the $J_{\mathrm{T}}$ from Eq.~\eqref{eq:Jt} into the Eq.~\eqref{eq:Sint} gives the Eq.~\eqref{eq:Sint2} of the main text.

\section{THE TRANVERSE CURRENT-CURRENT RESPONSE FUNCTION $\mathcal{R}_{\mathrm{T}}(\Omega,\vec{q})$}\label{sec:RT}
\subsection{The RPA equation}
The gauge field propagator is defined as
\begin{gather}
\mathcal{D}^{(0)}_{\mu\nu} = \langle T_{\tau} a_{\mu} a_{\nu} \rangle,
\end{gather}
where $T_{\tau}$ is the time ordering operator and $a_{\mu}$ is the $\mu$ th component gauge field. $\mu=\{0,1,2\}=\{\tau,x,y\}$. The average $\langle...\rangle$ is taken with respect to the bare gauge field action $S_{\mathrm{Coulomb}}$ defined in the main text. The superscript ``$(0)$" in $\mathcal{D}^{(0)}$ shows that it is a bare gauge field propagator. From the action $S_{\mathrm{CF}}$ the current operator is
\begin{gather}
J_{\mu}=\frac{\delta S_{\mathrm{CF}}}{\delta a_{\mu}}= v_F \bar{\psi} i\gamma_{\mu} \psi.
\end{gather}
Then the current-current response function can be defined as
\begin{gather}
\mathcal{R}_{\mu\nu}\equiv  \langle T_{\tau}  J_{\mu}  \, J_{\nu} \rangle =- \langle T_{\tau}  \bar{\psi}v_F \gamma_{\mu} \psi  \;  \bar{\psi} v_F\gamma_{\nu} \psi \rangle.
\end{gather}
Here the average is taken with respect to the mean field finite density Dirac Fermi sea. In defining $\mathcal{R}_{\mu\nu}$ we have only considered the paramagnetic contribution while ignored the diamagnetic contribution. However, this is justified. As shown in Ref.~\onlinecite{Principi2009}, for Dirac fermions the obital diamagnetic susceptibility is identically zero if the Fermi energy is not at the Dirac point. Fourier transformed into the momentum space the response function can be rewritten as
\begin{gather}
\mathcal{R}_{\mu\nu}(q)= v_F^2 \int \frac{d\omega d^2 \vec{k}}{(2\pi)^3}\mathrm{Tr}[\gamma_{\mu} \mathcal{G}(k)\gamma_{\nu} \mathcal{G}(k+q)],
\end{gather}
where $k\equiv(\omega,\vec{k})$ and $q=(\Omega,\vec{q})$. Remember that we have already taken the $T=0$ limit so that all Matsubara frequencies $\omega,\Omega$ are continuous. The trace ``$\mathrm{Tr}$" is taken with respect to the pseudo-spin indices. $\mathcal{G}(k)$ is the free Dirac CF propagator given by
\begin{align}
\mathcal{G}(k) & \equiv \langle T_{\tau} \psi(k) \bar{\psi}(-k) \rangle  = \frac{1}{\gamma_0 (i\omega -\epsilon_F) - i\gamma_j k_j }.
\end{align}

With these definitions of the gauge field propagator and response functions we can write down the RPA equation as
\begin{gather}
[\mathcal{D}^{\mathrm{RPA}}]^{-1}=[ \mathcal{D}^{(0)}]^{-1} -\mathcal{R},
\end{gather}
which is a matrix equation. However, since $\mathcal{D}^{(0)}$ has only a spatial transverse component, we only need to consider the transverse current-current response function: $\mathcal{R}_{\mathrm{T}}( \Omega,\vec{q})= [\delta_{ij}-\hat{q}_i \hat{q}_j] \, \mathcal{R}_{ij}(\Omega,\vec{q})$. The computation of $\mathcal{R}_{\mathrm{T}}(\Omega,\vec{q})$ has been done before and can be found in for example Ref.~\onlinecite{Miransky2001}. The final result, written in terms of our notations, is
\begin{align}
\mathcal{R}_{\mathrm{T}}(\Omega,\vec{q}) & = \frac{\epsilon_F}{2\pi} [-\frac{\Omega^2}{v_F^2|\vec{q}|^2}+\sqrt{1+\frac{\Omega^2}{v_F^2|\vec{q}|^2}}\frac{|\Omega|}{v_F|\vec{q}|}] \\
  &\approx \frac{\epsilon_F}{2\pi }\frac{|\Omega|}{v_F|\vec{q}|},
\end{align}
where in obtaining the last expression we have considered the limit $|\Omega|<v_F |\vec{q}|\ll \epsilon_F$.



%

\end{document}